\documentclass[%
reprint,
showpacs,
 amsmath,amssymb,
 aps,
prstab,
floatfix,
]{revtex4-1}

\usepackage{graphicx}
\usepackage{dcolumn}
\usepackage{bm}

\begin{document}

\newcommand{\D}{\,\mathrm{d}}

\title{Electron spectra and coherence of radiation in undulators}

\author{Eugene Bulyak}
\email{bulyak@kipt.kharkov.ua}
\affiliation{National Science Center `Kharkov Institute of Physics and Technology', 1 Academicheskaya str, Kharkov, Ukraine}

\author{Nikolay Shul'ga}
 \affiliation{National Science Center `Kharkov Institute of Physics and Technology', 1 Academicheskaya str, Kharkov, Ukraine}
\affiliation{V. N. Karazin National University, 4 Svodody sq., Kharkov, Ukraine}
\date{\today}

\begin{abstract}
Most bright sources of the radiation in hard x-ray and gamma--ray regions are undulator sources and Compton based ones. These sources are ultimate for production of polarized positrons necessary for future linear colliders ILC, CLIC.
We developed a novel method for evaluating the energy spectrum of electrons emitting the undulator- and the inverse Compton radiation. The method based on Poisson weighted superposition of electron states is applicable for whole range of the emission intensity per electron pass through the driving force, from much less than unity emitted photons (Compton sources) to many photons emitted (undulators), and for any energy of the photons. The method allows for account contributions in the energy spread both from the Poisson statistics and diffusion due to recoils.  The theoretical results were confirmed by simulations.  The electron energy spectrum was used for evaluation of the on-axis density of photons and their coherency making use of the `carrier--envelope' presentation for the emitting photons.  The evaluated maximum coherency degree of single--electron radiation is evaluated to be proportional to the undulator spatial period and inversely to the energy of electrons, the number of coherently emitted undulator periods almost independent of the undulator deflection parameter. The results of our study are applicable both for the classical limit of classical undulator and for the quantum limit of Compton gamma--ray sources.
\end{abstract}


\maketitle

\section{Introduction}
Most bright sources of the radiation in hard x-ray and gamma--ray regions are undulator sources and Compton based ones. These sources are ultimate for production of polarized positrons necessary for future linear colliders \cite{ilctdr,clicdr}.

Both of the processes -- undulator radiation and Compton inverse radiation -- exploit the same physical principle (radiation emission by electrons moving along sinusoidal or helical trajectory), and are known and theoretically described a few decades ago, see, e.g.  \cite{kim88,shulga96,hofman04}.

Classical description of the undulator radiation predicts maximal brightness attained on the undulator axis proportional to squared number of the periods of undulator magnetic structure, $\propto N_{\text{u}}^2$ due to spatial interference of the waves emitting by the electron.

With increasing the electron energy and/or shortening the undulator periods, the energy of undulator radiation advances to hard x-ray or gamma-ray region where the quantum effects in radiation emission will be expected. The undulator radiation would resemble the inverse Compton radiation, brightness is proportional to number of the laser photons with which the electron interacts, $\propto N_{\text{las}}$ (no interference of individual photons is expected).

In \cite{geloni12} there was emphasized that under the quantum approach, instead of number of the undulator periods one should account for number of photons emitted, which actually much smaller than the number of periods. Nevertheless, since number of photons emitted directly proportional to number of periods, the result of \cite{geloni12} yields the same quadratic dependence of brightness as in the classical approach.

In other words, these two approaches, the quantum one based on the Klein--Nishina electron--photon interactions, see \cite{klein}, and the classical approach yield different spectral--angular distribution of radiation. The quantum approach implies statistically independent quanta emitted, while under the classical approach all the quanta considered  interfering with each other.

We studied limitation of coherency of undulator radiation caused by recoils undergone by ultrarelativistic electrons emitting the photons. Due to the recoils, the energy of the electron is degraded and acquired a finite spread (due to a statistical nature of emission of the individual quanta). Under the classical approach, the undulator radiation is suggested either to keep the electron energy constant \cite{kincaid77,howells92}, or to adiabatically decrease (`chirping effect' that can be eliminated by means of tapering the undulator, see \cite{bosco83}).

Papers on quantum approach in undulator radiation predicted different energy distributions: in \cite{robb11} there was obtained a brush--like distribution depending on the number of the undulator periods, in \cite{geloni12} the distribution is continuous depending on number of quanta radiated out. In a recently published paper \cite{agapov14}, the energy spectrum of electrons moving along an undulator is evaluated within the diffusion approximation. The energy distribution is continuous, some fraction of the initial delta ensemble got acceleration (the authors commented it as somewhat nonphysical). A case of multiple Compton backscattering considered in \cite{kolchuzhkin2003} shows diffusive behavior of the electron spectra.

The present paper is intended to cover the gap between a classical undulator radiation of relatively low--energy electrons  and a quantum one of high--energy electrons and the inverse Compton radiation (characterized by small number of quanta emitted per the pass through laser pulse).

The paper is structured as follows. In the first section there described is a method to compute the energy spectrum of electrons passing along the undulator axis. The method is applicable for any number and any amplitude of recoils. The second section comprises results of study on coherency of photons emitted along the axis by electrons with statistical parameters evaluated in the first section. A method of `carrier--envelope' presentation is employed, which allows for estimation of the degree of coherency. A criterion for coherency degree is proposed. The third concluding section summarises the results and discusses possibility to attain coherency and thus high density of radiation in gamma--ray sources.

Numerical examples explore the ILC baseline source of polarized positrons: energy of electrons 150\,GeV, a helical undulator with the period 11.5\,mm, active length 127\,m and the undulator parameter $K=0.92$ (see \cite{ilctdr}), if other values are not indicated.

\section{\label{sec1}Electron spectrum along the undulator}
\subsection{General method}
Consider a certain pass length along the undulator (from the entrance to the coordinate $z$), in which an electron in average emits $\xi (z)$ photons:
\[
\xi\equiv \frac{\text{average energy loss } }{\text{average energy of photon}}\; .
\]
The average number of emitted photons is increasing linearly with $z$ from $\xi(0) = 0 $ up to $\xi(L_u) = \xi_* $ at the undulator exit.

Number of emitted photons is supposed to obey the Poisson distribution law as being the statistically independent rare events (see e.g. \cite{feller57,vankampen98,artru14}):
\begin{equation}\label{eq:poison}
f_n(\xi)=\frac{\xi^n\mathrm{e}^{-\xi}}{\Gamma (n+1)}\; ,
\end{equation}
where $n=1,2,\dots$ is the  number of photons, $\Gamma (n)$ the gamma function.

\begin{figure} 
\includegraphics[width=\columnwidth]{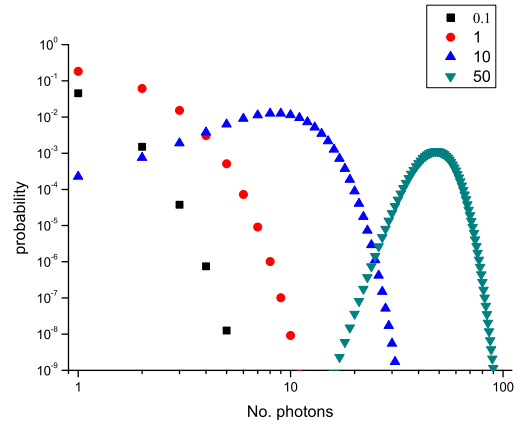}%
\caption{\label{ProbPhot}Expected number of emitted photons for $\xi = 0.1,1,10,50$.}
\end{figure}

Figure~\ref{ProbPhot} presents probability of emitting $n$ photons by the electron at different \emph{average} number of photons $\xi $.

Eq.~\eqref{eq:poison} yields that the fraction of particles has not emitted a photon exponentially decreased  with pass length:
\[
f_0 (\xi)=\mathrm{e}^{-\xi}\; .
\]

A vector of photon number probabilities (weights) can be defined with the components given by Eq.~\eqref{eq:poison}:
\begin{equation} \label{eq:stnumbers}
\vec{f} = \{ f_n(\xi) \}\; .
\end{equation}

\subsection{Spectrum of electron emitted $n$ photons}
Energy spectrum of an electron passing along an undulator can be presented as composed from the partial spectra corresponding to recoils caused by emission of $n$ photons. We attribute the $n$-th electron state as average energy spectrum, normalized to unity, due to emission of $n$ photons. Then the vector of states (in Hilbert space) can be constituted as:
\begin{align} \label{eq:states}
\vec{F} &= \{ F_n(\gamma ;\gamma_0,K,\lambda_u) \}\;
\intertext{with normalization}
&\int_1^\infty F_n(\gamma )\D\gamma = 1\; , \nonumber
\end{align}
where $\gamma_0$ is the initial energy (Lorentz factor) of the electron at undulator's entrance, $K,\lambda_\text{u} $ the undulator parameter and period, respectively.

Emission of a photon with energy $ \hbar\omega $  decreases the energy of electron:
\begin{equation} \label{eq:downshift}
\gamma' = \gamma - \frac{\hbar\omega}{ mc^2} = \gamma - \gamma_\text{ph}\; .
\end{equation}
It is induced spread in energy as well.

Let $W(\gamma_\text{ph}, \gamma)$ be a normalized to unity spectrum of emitted photons -- i.e. spectrum of recoils -- (here and below we omit indication of the undulator parameters $\lambda_\text{u},K$). Then, in general,  the components of the vector \eqref{eq:states} can be sequentially calculated:
\begin{equation} \label{eq:spectn}
F_{n}(\gamma) = \int_{\gamma_\text{ph}} F_{n-1}(\gamma+\gamma_\text{ph})W(\gamma_\text{ph}, \gamma+\gamma_\text{ph})\D\gamma_p\; ,
\end{equation}
starting with $F_{0}(\gamma ) =\delta (\gamma -\gamma_0)$ ($\delta $ is the Dirac delta--function).

Making use of centered the recoil spectra around the average value $\gamma_\text{ph} = \epsilon + b$ with $b(\gamma )\equiv <\gamma_\text{ph}>$, Eq.~\eqref{eq:spectn} may be reduced to
\begin{equation} \label{eq:spectnc}
F_{n}(\gamma - b) = \int_{\epsilon }  F_{n-1}(\gamma+\epsilon)w(\epsilon,\gamma) \D \epsilon \; .
\end{equation}
Thus $n$-th state is translation by $b$ the convolution of $n-1$ state with the recoil spectrum.

The aggregate spectrum of the electron having emitted \emph{in average} $\xi $ photons is dot product of the state vector \eqref{eq:spectn} and the vector of `weights' \eqref{eq:stnumbers}
\begin{equation} \label{eq:genspectr}
\mathcal{F}(\gamma,\xi ) = \vec{F}\cdot \vec{f} = \sum_{n=0} F_{n}(\gamma) \frac{\xi^n\mathrm{e}^{-\xi}}{\Gamma (n+1)} \; .
\end{equation}

For the undulator and Compton cases the recoil is small, the spectrum of recoil $w(\epsilon)$ is almost independent of $\gamma $: $b/\gamma \approx b/\gamma_0 \ll 1$. For this case $b(\gamma) = b(\gamma_0)$ is constant, the average energy of the $n$ state and the variance are
\begin{align*}
<\gamma>_n &= \gamma_0 - nb\; ; \\
\sigma_n^2 &\equiv <(\gamma +nb -\gamma_0)^2> = n \sigma_1^2\; .
\end{align*}

Making use of these simplifications, the first two moments of the aggregate spectrum read:
\begin{align}
<\gamma >_\xi &= \sum f_n(\xi) \int\D\gamma F_n(\gamma) \gamma  = \gamma_0 - \xi b\; ; \\
<\gamma^2 >_\xi &=\sum f_n(\xi) \int\D\gamma F_n(\gamma) \gamma^2  \nonumber \\ &= \gamma_0^2 + \xi (\sigma_1^2-2\gamma_0 b) + \xi (1+\xi) b^2\; ; \\
\sigma_\xi^2 &\equiv  <\gamma^2> - <\gamma >^2 = \xi (\sigma_1^2+b^2)\; . \label{eq:cumspread}
\end{align}

\subsection{Estimations for undulators}
Let us consider, as an example, application of the developed method to the case of `weak' undulator, which is also equivalent to the Compton backscattering source.
The spectral density of the undulator photons in the fundamental harmonic is (see e.g. \cite{shulga96}):
\begin{equation} \label{eq:spectr1}
S(\zeta, K) = G(K)\left[ 1-2\zeta(1-\zeta)\right]\mathrm{\Theta}[1-\zeta (1+K ^2)]\; ,
\end{equation}
where $G(K)$ is the normalizing factor, $\mathrm{\Theta}(x)$ represents the Heaviside theta function; $\zeta $ is a reduced energy of photons; $\zeta \equiv \epsilon_{\text{phot}}/(2\gamma^2\hbar\omega_{\text{u}})$.

For the case of `weak' undulator, $K^2\ll 1; \; G(K)=3/2$, the centered spectrum of recoils reads:
\begin{equation} \label{eq:centrecoil}
w(\epsilon,\gamma) = \frac{3}{8 b^3}(b^2+\epsilon^2)\Theta(b+\epsilon)\Theta(b-\epsilon)
\end{equation}
where $b = \gamma^2_0 \lambda_{\text{C}}/\lambda_{\text{u}}$ with $\lambda_{\text{C}}$ being the Compton wavelength, $\lambda_{\text{u}}$ the undulator period.

\begin{figure}
\includegraphics[width=\columnwidth]{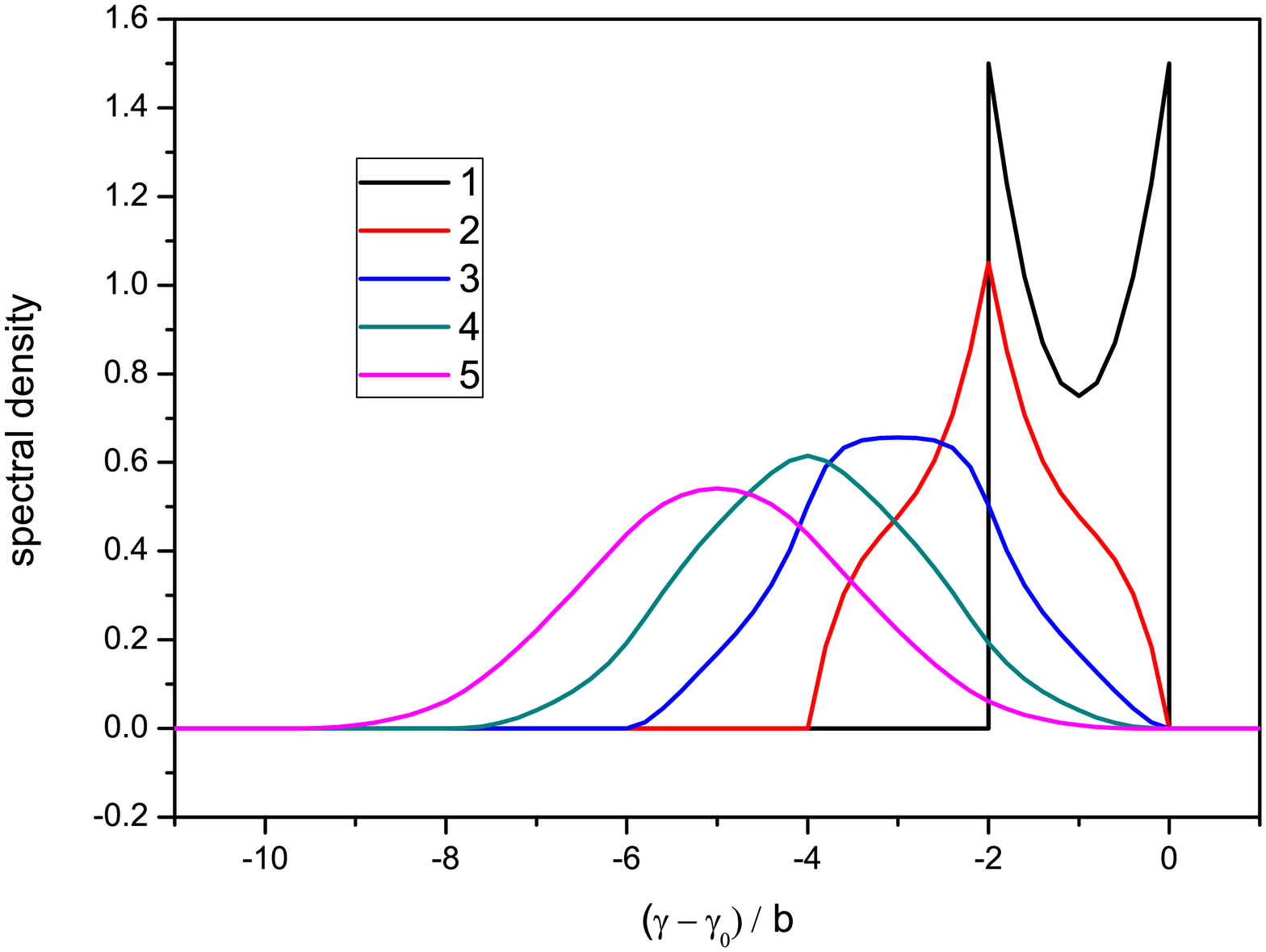}
\caption{\label{estates} The first five analytical electron states, $K^2 \ll 1$.}
\end{figure}

Substituting \eqref{eq:centrecoil} into \eqref{eq:spectnc}, we can successively derive the the electron states. The first five states are drawn in Fig.~\ref{estates}.

The electron states possess following specific characteristics:
\begin{itemize}
\item The average energy is $<\gamma >_n = \gamma_0 - n b$.

\item The base span is $(<\gamma >_n - nb) \le \gamma \le (<\gamma >_n + nb)$, no density above $\gamma_0$ and below $\gamma_0 - 2 n b$.

\item The variance of $n$-state density is $\sigma^2_n = n\sigma_1^2$ with $\sigma_1^2=\frac{2}{5}b^2$.

\item With increase of $n$ the density shape gradually loses its individuality and tends to the Gaussian shape according to the Central Limit Theorem.
\end{itemize}

It should be emphasized that the major contribution to the cumulative energy spread \eqref{eq:cumspread} comes from Poisson superposition of the states $\xi b^2$ as compared with the width of the spectrum $\sigma_1^2$:
\begin{equation} \label{eq:cumspreadu}
\sigma_\text{e}^2(\xi )  = \xi (\sigma_1^2+b^2) = \xi b^2 (1+2/5)\; .
\end{equation}

It also should be mentioned that for 1D model considered in \cite{robb11} (suggested a spectrum of emitted photons being of a delta--function shape, $W(\gamma_p, \gamma) = \delta(\gamma_p-\gamma_p^\text{max})$ ) the spectrum of electrons derived by means of our method coincides with that obtained by authors of \cite{robb11}: narrow picks distributed according to the Poisson law.

\begin{figure}
\includegraphics[width=\columnwidth]{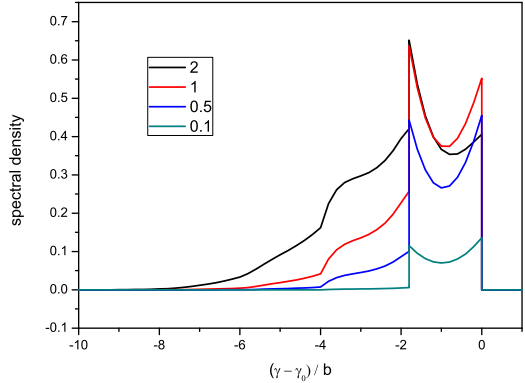}
\caption{\label{cumspec012} Cumulative spectra for the different $\xi$'s. Superposition of the first 5 states.}
\end{figure}

\begin{figure}
\includegraphics[width=\columnwidth]{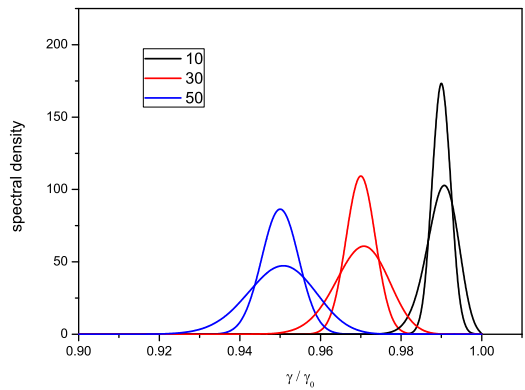}
\caption{\label{cumspec1050} The central component, $n=\xi$, (higher), and the cumulative spectra (wider) for $\xi= 10,30,50$; $b=0.001 \gamma_0$. }
\end{figure}

Cumulative spectra of the first 5 states for relatively small $\xi$ are presented in Fig.~\ref{cumspec012}, for large $\xi$  (all the states approximated by the Gaussian functions) in Fig.~\ref{cumspec1050}.

\begin{figure*}
\includegraphics[width=0.9\columnwidth]{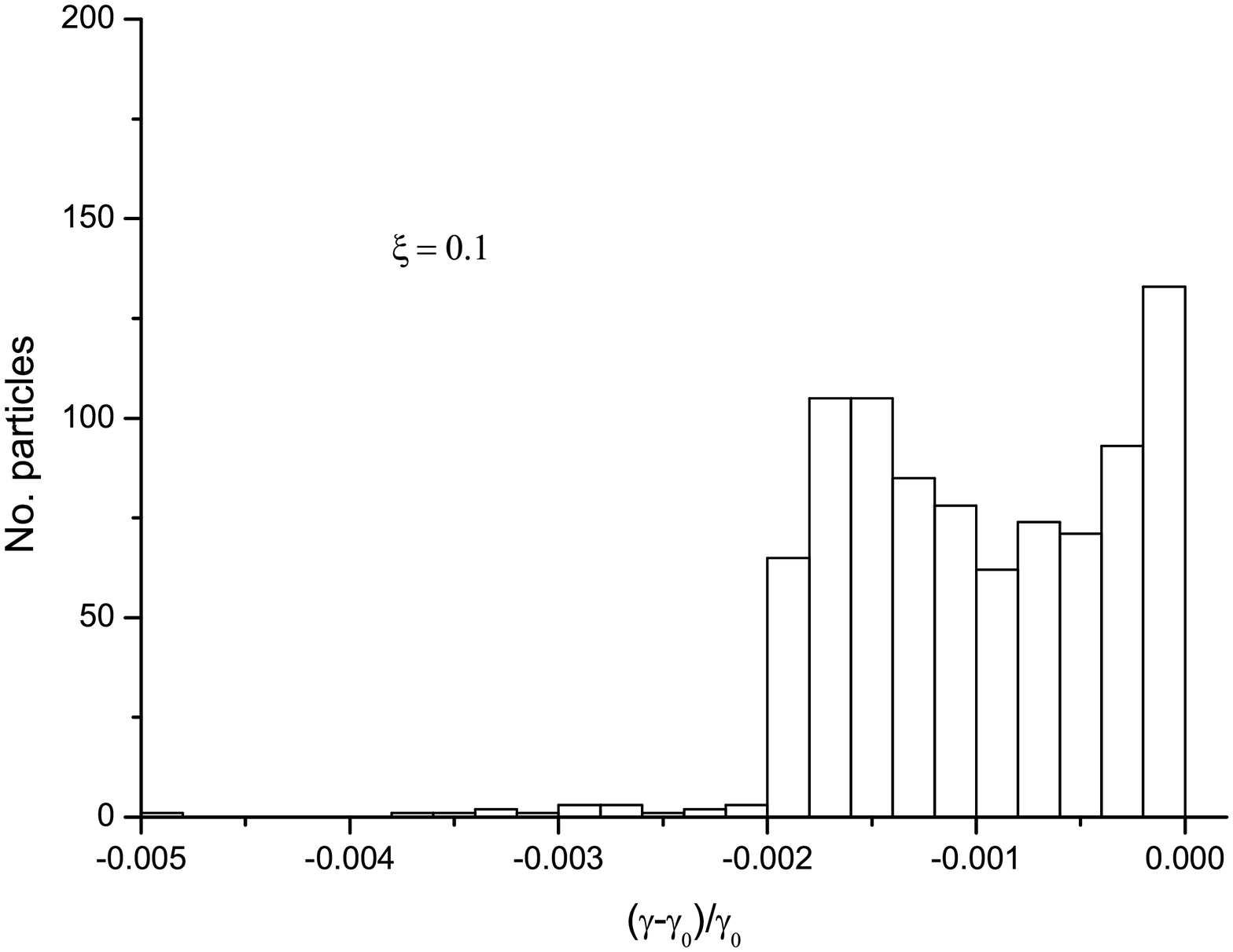}
\includegraphics[width=0.9\columnwidth]{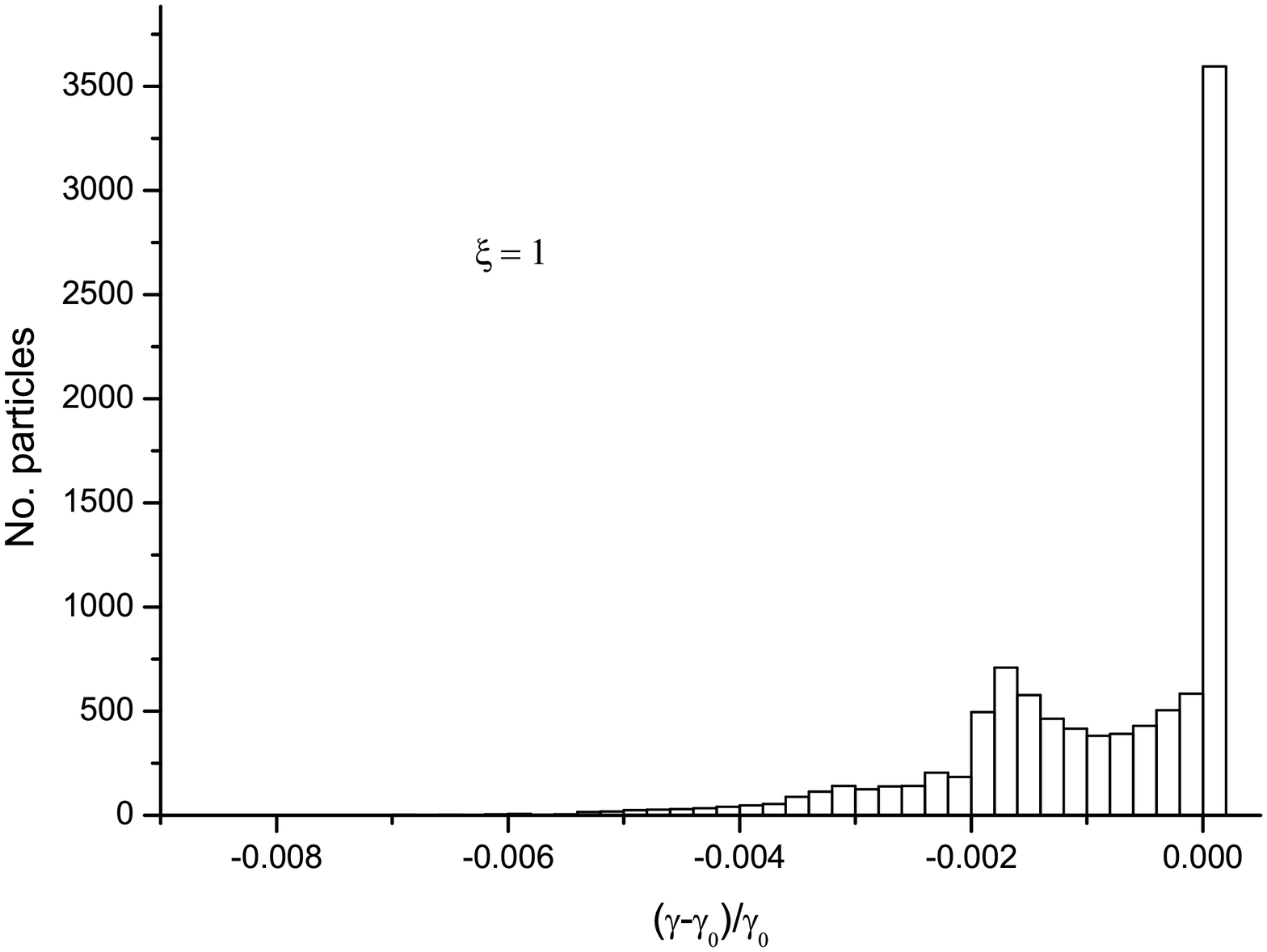}\\
\includegraphics[width=0.9\columnwidth]{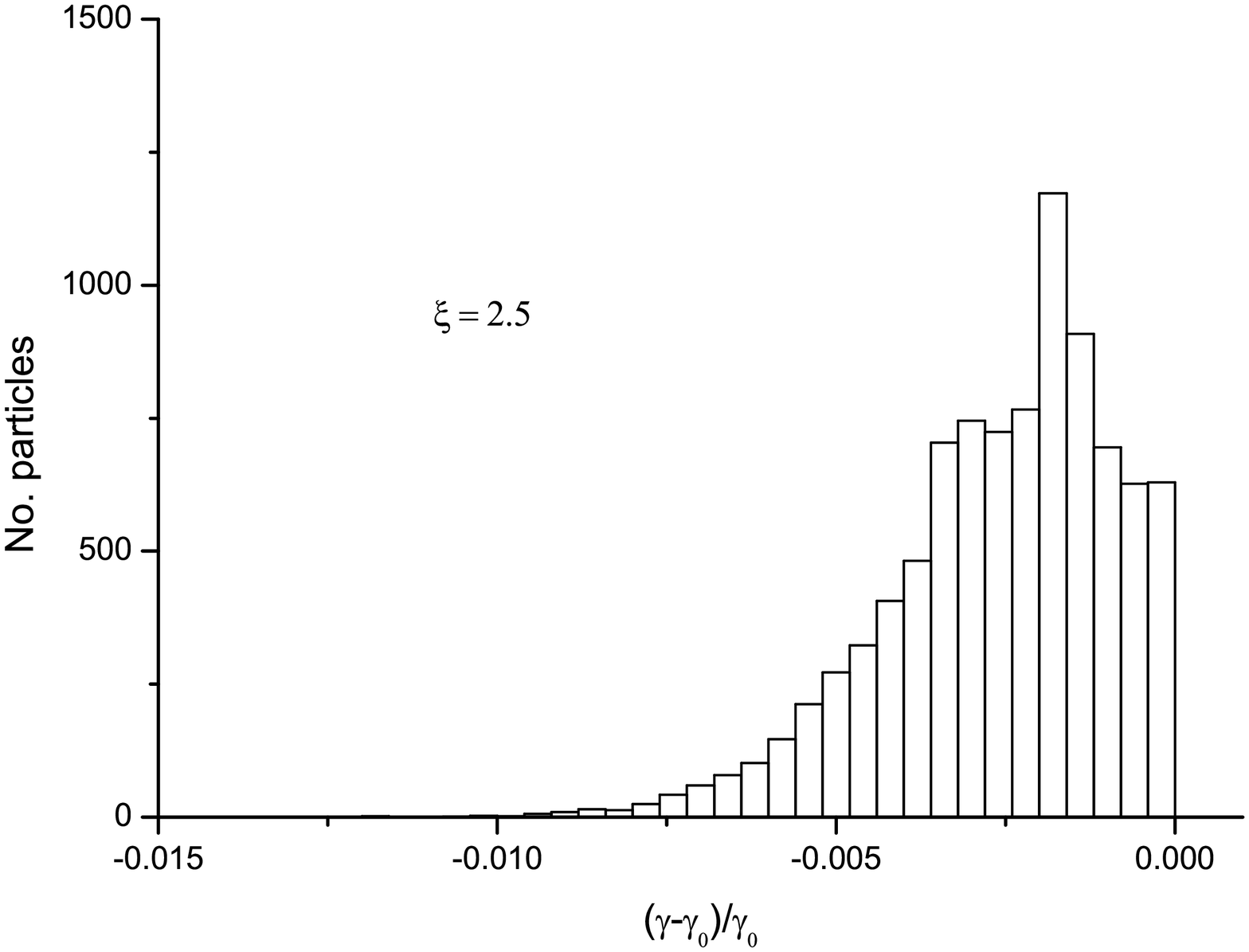}
\includegraphics[width=0.9\columnwidth]{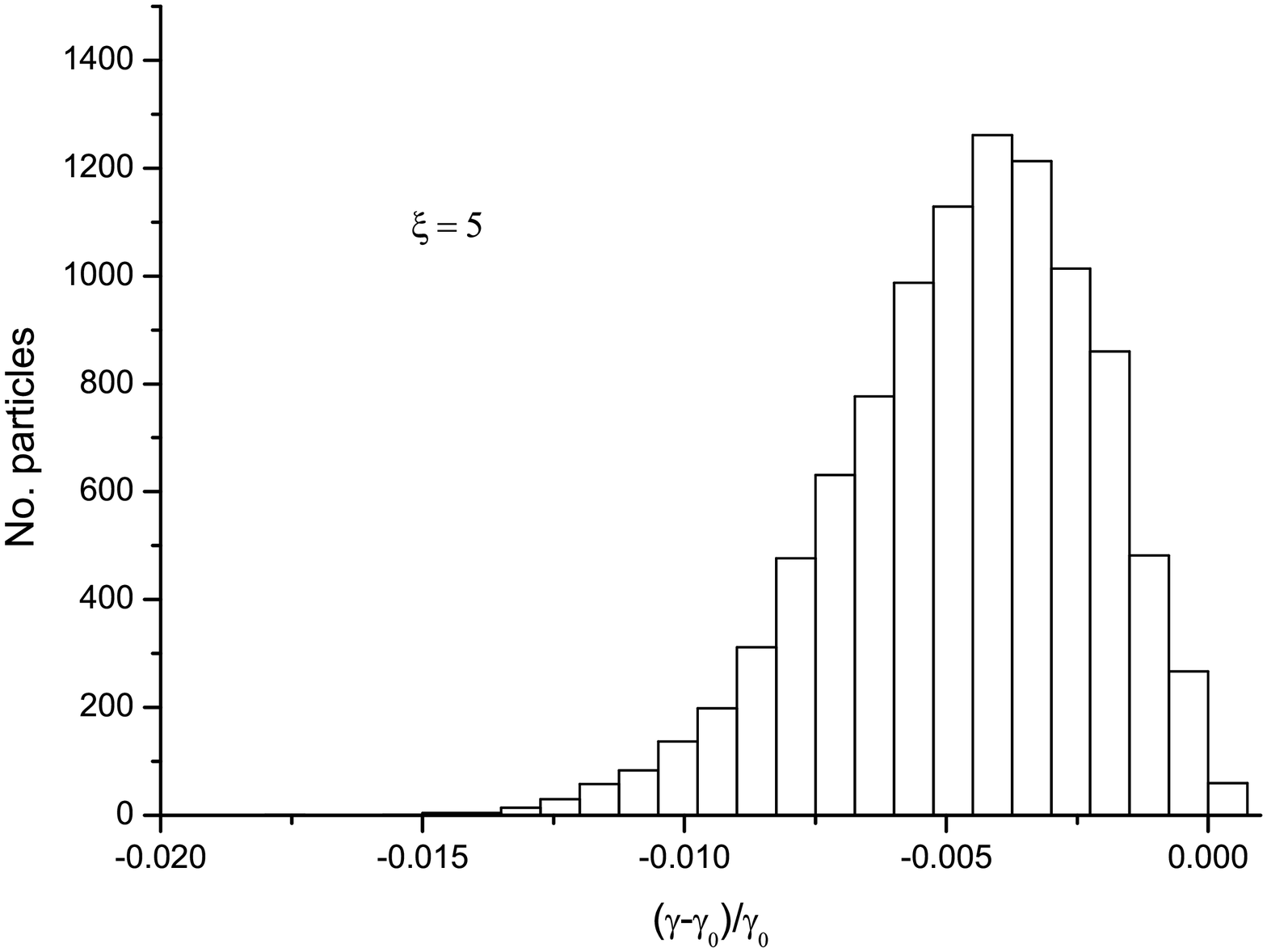}\\
\includegraphics[width=0.9\columnwidth]{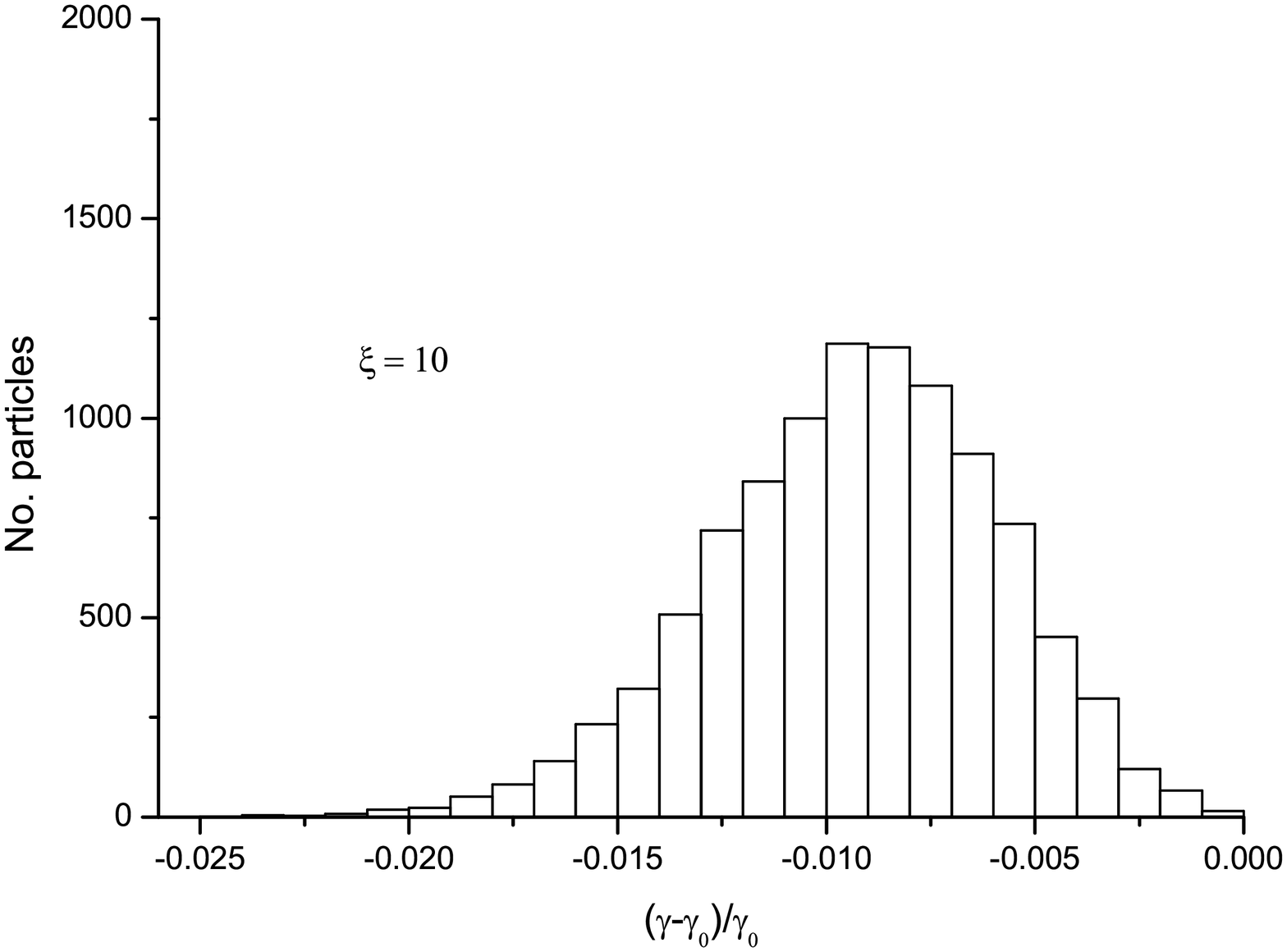}
\includegraphics[width=0.9\columnwidth]{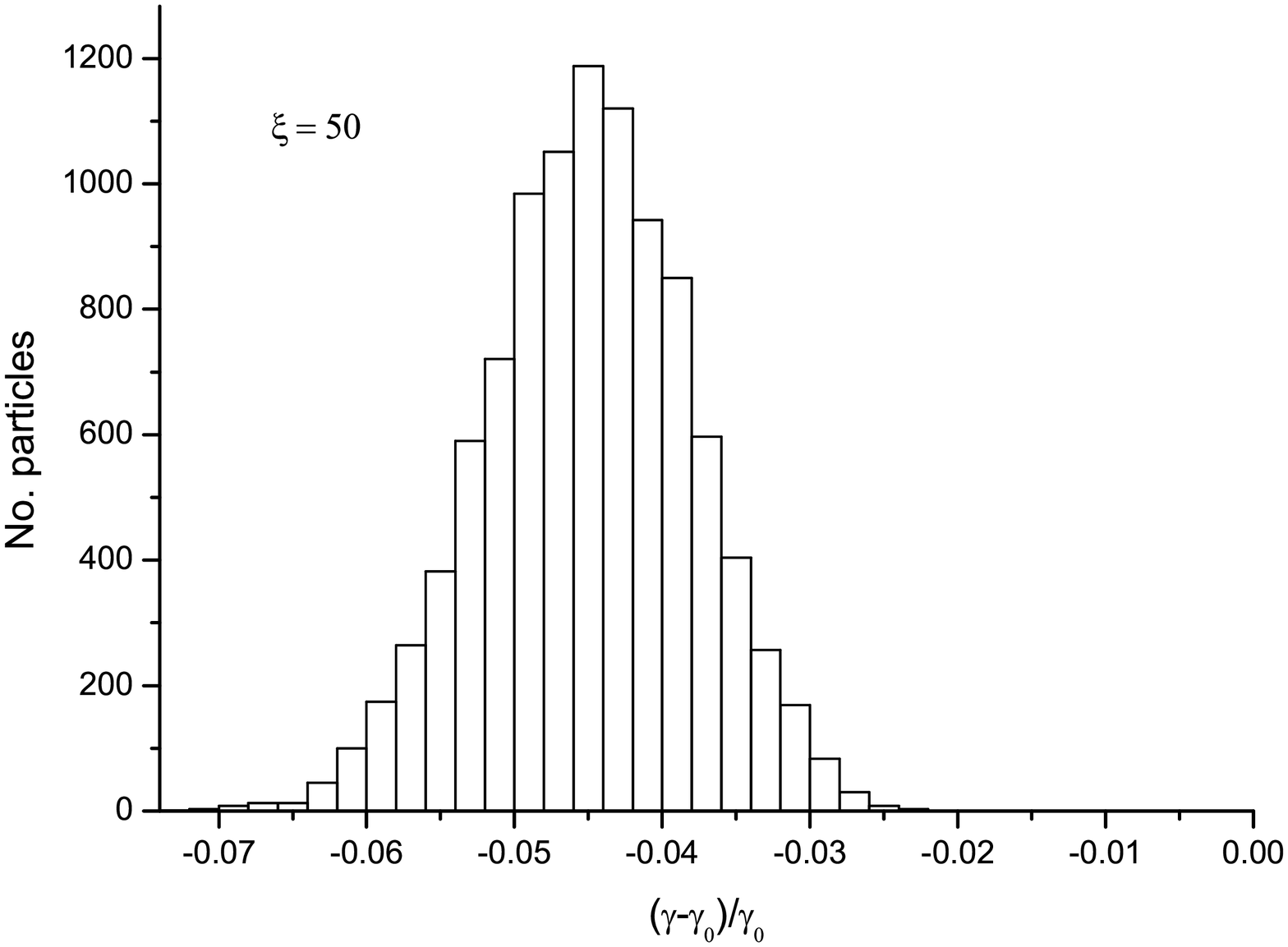}
\caption{\label{fig:histo} Simulated spectra of electrons.  For $\xi = 0.1$ -- top left -- the non--recoiled particles are not displayed.}
\end{figure*}

In order to validate the analytical estimations, a Monte-Carlo code has been used to simulate the process of undulator radiation. The code, mainly intended for simulation performance of the Compton sources of hard radiation (see, e.g. \cite{bulyak14a}), simulates the recoils of electrons due to a random emission of individual quanta with the spectrum \eqref{eq:spectr1}. Fig.~\ref{fig:histo} represents spectra of electrons from simulations. Total number of particles 10000, the maximum number of quanta 800, the maximum energy of photons $\epsilon_* = 0.002 E_\text{init}$, the deflection parameter $K\ll 1$.

As it is seen from the figure, the recoil spectrum resemble the fundamental harmonic with additional low-energy tail of small intensity for $\xi\lesssim 1$. For the practical cases, $\xi\gg 1$ the energy spectra resemble the normal distribution.

\subsection{Number of emitted photons}
Based on a brief description of the undulator radiation \cite{kincaid77,xray}, we estimate the length along an undulator axis to emit one photon in average by the electron.

The total energy (in average) radiated by a single electron traversing the $N_{\text{u}}$--period undulator is given by
\begin{equation}\label{eq:totenergy}
p_{\text{t}} = \frac{4}{3}\pi\alpha\hbar N_{\text{u}} \omega_{\text{u}}\gamma^2 K^2\;,
\end{equation}
where $N_{\text{u}}$ is the number of undulator periods, $\omega_{\text{u}}\equiv 2\pi c/\lambda_{\text{u}})$ the undulator equivalent frequency, $\lambda_{\text{u}}$ the undulator period, $K$ the the deflection parameter, $\alpha$ the fine structure constant. We suggest a helical undulator, for a planar one the energy is half of \eqref{eq:totenergy}.


The average energy of the undulator photon is almost independent of the undulator parameter $K$ and equal to half of maximum energy in the fundamental harmonic at $K\to 0$, see \cite{hofman04}.

Substituting the average recoil energy $b$ into \eqref{eq:totenergy}, we can derive the number of undulator's periods $\nu_1$ to emit one photon by the electron:
\begin{equation}\label{eq:numundper}
\nu_1 \approx \frac{3}{4\pi\alpha K^2}\;.
\end{equation}

For the case of $K =1$, we have
\[
\nu_1^\text{helical}\approx 137/4 \approx 17\,\text{periods}\;.
\]
Accordingly, for the planar undulator $\nu_1^\text{planar}\approx 137/3\approx 46\;\text{periods}$. This estimation is in agreement with \cite{howells92} where $\nu_1^\text{planar}\sim 1/\alpha\;\text{periods}$.

Transition to the planar undulator corresponds to replacement $K\to K/\sqrt{2}$ in \eqref{eq:spectr1}. For the Compton sources, the undulator spatial period should be substituted by $\lambda_\text{u} \to \lambda_\text{las}/2$ (for the head--on crossing of the electron with the laser pulse, see e.g. \cite{bulyak05}).

\section{\label{sec2} Frequency spread and coherence}
\subsection{On-axis density and coherence}
In the preceding section we regularized the electron kinetics: the energy spectrum of the electron passing along an undulator was derived as function of its position, initial energy, and the undulator parameters: the field strength and  the period. In the present section, we use two first moments of the spectrum, the average energy and the spread (r.m.s. width of the spectrum), to evaluate a coherence degree of radiation emitting by the electron. The radiation will be presented as $\xi_L$ individual pulses, uniformly distributed along the undulator, the distance between consecutive pulses is equal to $L_{\text{u}}/\xi_L$ ($L_{\text{u}}$ is the undulator length). We will consider coherence of the photons in the fundamental harmonic emitting along the undulator axis -- 1D model.

The degree of coherence will be estimated with the following reasoning.
The pulses are considered coherent if their field strengths are piled up, incoherent if the energies of the pulses are summed. The density of energy provided by $j$ pulses reads
\begin{equation} \label{eq:densdef}
\mathcal{D}(j) = \int\limits_{-\infty}^\infty \left( \sum _{n=1}^j u_n(\tau-\tau_n) \right)^2 \D \tau\; ,
\end{equation}
where $u_n(\tau)$ is the electrical field strength of $n$-th pulse, $\tau_n $ is the offset.

If  the pulses are  normalized quadratically to unity,
\[
\int u_n^2(\tau )\D\tau = 1\; ,
\]
then $j$ pulses are considered fully coherent if power density is proportional to $j^2$, and fully incoherent if  proportional to $j$.
A degree of coherency may be attributed as (see, e.g. \cite{klauder68,born70}):
\begin{equation} \label{eq:cohdef}
\Gamma(j ) = \frac{\ln \mathcal{D}(j)}{\ln j} -1\; .
\end{equation}

The degree of coherency \eqref{eq:cohdef} varies from $\Gamma = 1$ -- fully coherent pulses, energy density $\propto j^2$ -- to $\Gamma_j = 0$ -- fully incoherent, the maximum density $\propto j^1$. As it follows from \eqref{eq:densdef}, the coherency is determined by the terms $\gamma_{m,n} = \int u_m(\tau-\tau_m) u_n(\tau-\tau_n) \D \tau$. Full coherency is equivalent to $\gamma_{m,n} = 1$, incoherency to $\gamma_{m,n} = \delta_{mn}$ ($\delta_{mn}$ is the Kronecker delta).

\subsection{Regularization of emitted pulses}
The pulses generated by the electron passing a long undulator, $L_u/\lambda_u \gg 1 $, may be represented with the ``carrier--envelope'' signal, see e.g. \cite{brabec}, where the carrier frequency corresponds to the average energy of the electron, the pulse length is inversely proportional to the frequency spread induced by the spread of electron energy.

For a general form of a pulse,
\begin{equation} \label{eq:uxtgen}
u(x,t) = \frac{1}{\sqrt{2\pi}}\int_{-\infty}^{\infty} A(k)\exp[- i (\omega(k) t - k x) ]\D k\; ,
\end{equation}
where $x$ is scaled along the undulator axis, the photon pulse spectrum, $A(k)$ is related to the pulse envelope $u(x,0)$ via the Fourier transform \eqref{eq:agen}, and vice versa.

In a dispersion--free space, $\omega (k) = k c$, it reads:
\begin{equation} \label{eq:agen}
A(k) = \frac{1}{\sqrt{2\pi}}\int_{-\infty}^{\infty} u(x,0) \exp[-i  k x ]\D x\, .
\end{equation}
At these conditions a pulse preserves its form and travels at the speed of light $c$.

Since we suggested that the frequency spread in radiation is induced by the energy spread of the electrons, and the former resembles the normal distribution, 
we will consider gaussian pulses (packages).
Introducing a normalizing factor $P$ such that $\int u^2(x,t) \D (x-ct) = 1$, and substituting it into \eqref{eq:agen}, we get
\begin{align} \label{eq:ufinal}
u(\tau ; k_0,\sigma ) &=\left[ \frac{2}{\sigma\sqrt{\pi } \left[ 1+\exp (-k_0^2\sigma ^2)\right] }\right] ^\frac{1}{2} \times \nonumber \\
&\exp \left( -\frac{\tau ^2}{2\sigma ^2}\right)\cos\left(k_0\tau \right)\; ,
\end{align}
with $\tau\equiv x - c t$ and $k_0 = 1/\lambdabar_\text{carrier}$.

The interference term of two gaussian pulses,
\[
R(x; k, \sigma) = H \exp \left( - \frac{x^2}{2\sigma^2}\right) \cos (k x)\; ,
\]
with
\[
H(\sigma, k) = \sqrt{\frac{2}{\sqrt{\pi} \sigma \left[1+\exp(-\sigma_2^2 k_2^2)\right]}}
\]
being the norm of a pulse, offset by $\tau $, possesses an exact form
\begin{widetext}
\begin{eqnarray} \label{eq:exint}
\gamma_{m,n} &\equiv& \int_{-\infty}^\infty  R(x; k_m,\sigma_m) R(x-\tau ; k_n,\sigma_n) \D x =
\sqrt{\frac{2 \sigma_m \sigma_n}{(\sigma_m^2+\sigma_n^2)\left[1+\exp(-\sigma_m^2k_m^2)\right] \left[1+\exp(-\sigma_n^2k_n^2)\right]}}\nonumber \\
&\times& \left\{ \mathrm{e}^{\left[ -\frac{\sigma_m^2\sigma_n^2(k_m-k_n)^2+\tau^2}{2(\sigma_m^2+\sigma_n^2)} \right]} \cos\left[\tau\frac{k_m\sigma_m^2+k_n\sigma_n^2}{\sigma_m^2 + \sigma_n^2} \right] + \mathrm{e}^{\left[ -\frac{\sigma_m^2\sigma_n^2(k_m+k_n)^2+\tau^2}{2(\sigma_m^2+\sigma_n^2)} \right]} \cos\left[\tau\frac{k_m\sigma_m^2-k_n\sigma_n^2}{\sigma_m^2 + \sigma_n^2} \right]\right\} \; .
\end{eqnarray}

In practical cases a photon pulse in undulators comprises many periods, $\sigma k\lesssim N_{\text{u}}$. Therefore $\exp (-\sigma^2k^2) \sim \exp (- N_{\text{u}}^2)\lll 1$ and the interference term \eqref{eq:exint} can be simplified:
\begin{equation} \label{eq:exintapp}
\gamma_{m,n}\approx \sqrt{\frac{2 \sigma_m \sigma_n}{(\sigma_m^2+\sigma_n^2)}}\, \exp\left[ -\frac{\sigma_m^2\sigma_n^2(k_m-k_n)^2+\tau^2}{2(\sigma_m^2+\sigma_n^2)} \right]\, \cos\left[\tau\frac{k_m\sigma_m^2+k_n\sigma_n^2}{\sigma_m^2 + \sigma_n^2} \right] \; .
\end{equation}
\end{widetext}

Expression \eqref{eq:exintapp} allows for computation the power density \eqref{eq:densdef} and consecutively the coherence factor. The term \eqref{eq:densdef} consists of the exponential factor depending of widths of pulses, and an oscillating cosine factor responsible for variation of the carrier frequency between two pulses.

\subsection{Connection to the undulator radiation}
Within adopted approximation of small recoil, neglecting an initial energy spread in electrons and the frequency spread due to the maximal pulse width (equal to the reduced undulator length), the wave vector $k_0$ and the width of envelope cast into:
\begin{align*}
k_0 &= \frac{2\gamma^2}{\lambdabar_\text{u} (1+K^2)}\; ; \\
\sigma &=\frac{1}{\sigma_k} = \frac{\lambdabar_\text{u} (1+K^2)}{4\gamma\sigma_\text{e}}\;,
\end{align*}
where $\sigma_\text{e}$ is the dispersion of electron's energy, determined in Sect.~\ref{sec1}.

The reduction of the wave vectors and the envelope width, and the displacement $\tau$ read
\begin{align*}
k_m &= k_0 - m k_0^2\lambdabar_{\text{C}}/\gamma\; ;\\
\sigma^2_m &= \sigma_1^2/m = \frac{1}{m}\frac{5}{28 k_0^2}\left(\frac{\lambda_{\text{u}}}{\gamma\lambda_{\text{C}}}\right)^2\; ;\\
\tau_{n-m} &= (n-m)\frac{2\pi\nu_1}{k_0} = (n-m)\tau_1\; ,
\end{align*}

Ultimate coherency -- maximum on-axis density of radiation -- is attainable by means of keeping the carrier frequency constant. This can be done by proper tapering of the undulators, see \cite{bosco83,wang09,mun14}. Further we will refer to the cases of constant carrier frequency as `tapered'.

\begin{figure}
\includegraphics[width=0.9\columnwidth]{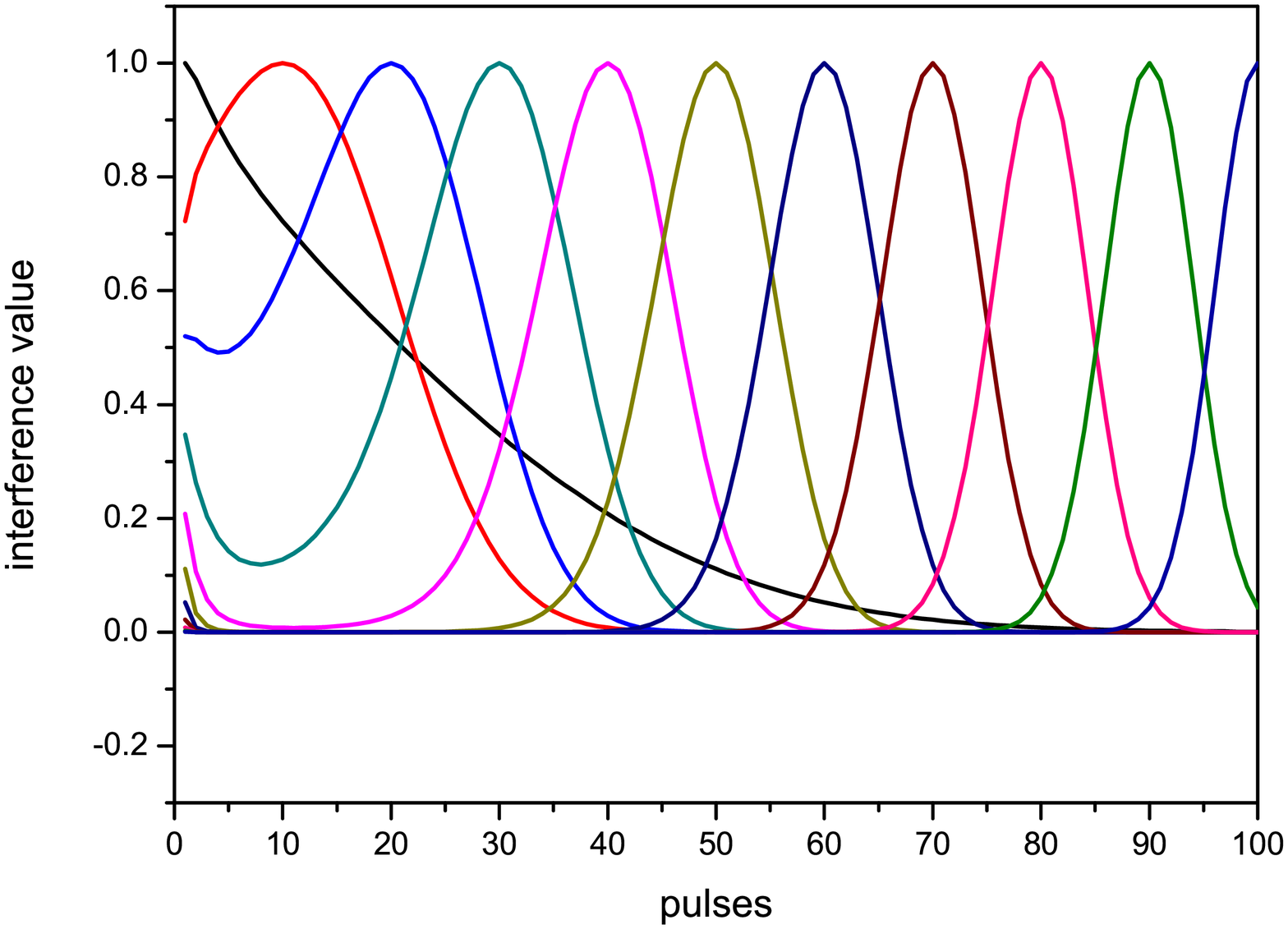}\\
\includegraphics[width=0.9\columnwidth]{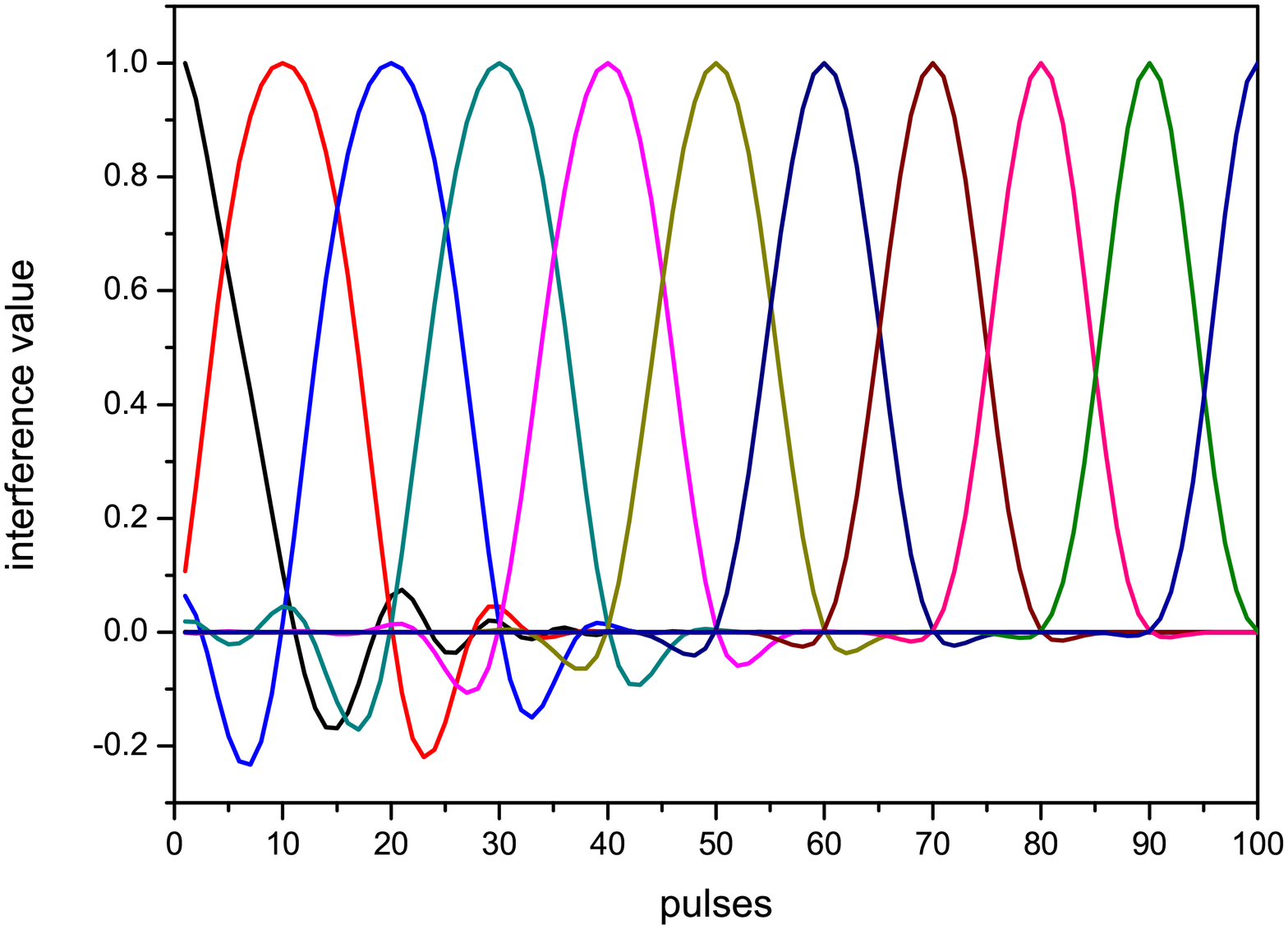}
\caption{\label{inter10} Interference terms magnitude $\gamma_{m,n}$ for different central pulse numbers $m=1,10,20,\dots , 100$, top -- `tapered' undulator, bottom -- untapered one.}
\end{figure}

Examples of the interference terms are plotted in Fig.~\ref{inter10}. As it can be seen from the plots, with increase in the pulse number $m$ interference with the neighboring pulses decreased faster. For a general (untapered) case, the interference terms undergone oscillations.

\begin{figure}
\includegraphics[width=\columnwidth]{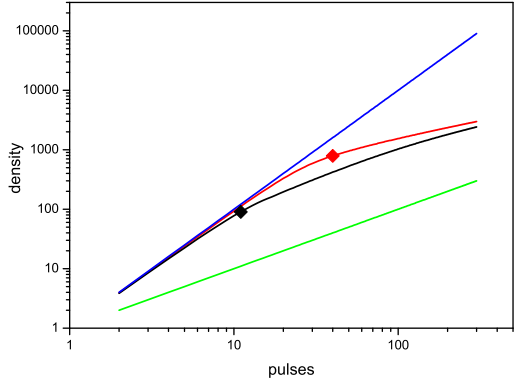}
\caption{\label{densvsphots} Density  vs. number of pulses; the blue curve represents full coherence, the green one is for no coherence, the red  for a `tapered' undulator, the black one is for a regular undulator. The diamonds indicate the coherence criteria. }
\end{figure}

\begin{figure}
\includegraphics[width=\columnwidth]{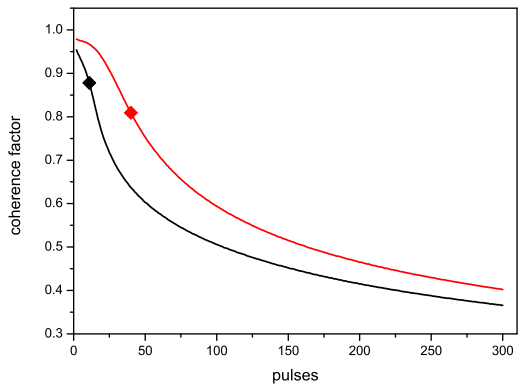}
\caption{\label{cohvsphot} Coherence factor vs. number of pulses, the black curve represents a regular undulator, the red one a `tapered'.}
\end{figure}

Energy density $\mathcal{D}_m$ as a function of number of emitted photons, beginning from the front end $m=1$ is plotted in Fig.~\ref{densvsphots}, the coherence factor in Fig.~\ref{cohvsphot}.

As it can be seen from Fig.~\ref{densvsphots}, radiation from the front end of the undulator displays coherency. With increasing in the number of radiated out pulses radiation becomes less coherent. Deviation from the coherent limit (the blue curve in Fig.~\ref{densvsphots}) is more severe for the non-tapered undulator.

\subsection{Coherency criterion}
A criterion that indicates coherency length extends of the undulator downstream, counted from the frontend (the most coherent radiation), may be established from a mere contraction that the exponent power in \eqref{eq:exintapp} is equal to minus unity:
\begin{equation} \label{eq:cohcrit}
\gamma_{1,\xi_*}\approx \exp(-1)\; \to \; \frac{\sigma_1^2\sigma_{\xi_*}^2(k_1-k_{\xi_*})^2+\tau^2}{2(\sigma_1^2+\sigma_{\xi_*}^2)}=1\; .
\end{equation}

For the most coherent tapered undulator, the number of coherent pulses is
\begin{equation} \label{eq:cohtaper}
\xi_*^{\text{(t)}} = \sqrt{\frac{5}{63}}\, \left(\frac{\alpha\lambda_{\text{u}}}{\gamma\lambda_{\text{C}}}\right)K^2\; .
\end{equation}

Multiplying $\xi_*$ by number of the undulator periods for emitting a single photon $\nu_1$, Eq.~\eqref{eq:numundper}, we get the number of the coherent undulator periods that turn out to be independent of the undulator parameter $K$:
\begin{equation} \label{eq:cohtapern}
N_{\text{coh}}^{\text{(t)}}\approx \frac{3}{4\pi}\sqrt{\frac{5}{63}} \left(\frac{\alpha\lambda_{\text{u}}}{\gamma\lambda_{\text{C}}}\right)\approx 2\times 10^{8} \frac{\lambda_{\text{u}}[\text{m}]}{\gamma}\; ,
\end{equation}
with $\lambda_{\text{u}}[\text{m}]$ being scaled in meters.

The case of untapered undulator is more complicated. Nevertheless, for a limiting case when the undulator parameter is not too small,
\begin{equation} \label{eq:cohcond}
\frac{K^2}{1+K^2} > 10^{-9}\frac{\gamma}{\lambda_{\text{u}}[\text{m}]}\; ,
\end{equation}
the number of coherent pulses is determined only by the undulator parameter:
\begin{equation} \label{eq:cohuntapern}
\xi_*^{\text{(ut)}}\approx 3 (1+K^2)^2\; .
\end{equation}

Positions of corresponding coherency criteria are indicated with diamonds in Figs.~\ref{densvsphots},\ref{cohvsphot}. As it is anticipated, the tapering substantially increases coherency of single--electron radiation.

\section{Summary and Conclusions}
A novel method for evaluating the energy spectrum of electrons emitting the undulator- and the inverse Compton radiation was developed. The method based on Poisson weighted superposition of electron states is applicable for whole range of the emission intensity per electron pass through the driving force, from much less than unity emitted photons (Compton sources) to many photons emitted (undulators). The method allows for account contributions in the energy spread both from the Poisson statistics and diffusion due to recoils. As it was shown, for the case of small number of photons emitted per pass the electron spectrum is mainly determined by the spectrum of recoils, while for many photons emitted major contribution comes from the Poisson law of recoils distribution. The theoretical results were confirmed by simulations. The evaluated width of electron spectrum, see \eqref{eq:cumspreadu}, is larger than that derived in \cite{robb11} for 1D Poisson model or the diffusion model considered in \cite{agapov14} since it accounts for the both processes.

The electron energy spectrum was used for evaluation of the on-axis density of photons and their coherency making use of the `carrier--envelope' presentation for the emitting photons.  The evaluated maximum coherency degree of single--electron radiation is obtained to be proportional to the undulator spatial period and inversely to the energy of electrons, the number of coherently emitted undulator periods almost independent of the undulator deflection parameter (for the ultimate case of the tapered undulators). The results of our study are applicable both for the classical limit of low-energy undulator radiation and for the quantum limit of Compton gamma--ray sources.

`Classical limit' for the coherency criterion corresponds to the case when all of photons are emitted at the resonant frequency and the pulse width (envelop) equal to the reduced length of an undulator, $\sigma _p = N_\text{u}\lambdabar _p (1+K^2)/(4\pi) $. The number of photons in the classical limit, $\hbar\to 0$, would be taken equal to the number of undulator periods. Thus we have the classical density $\mathcal{D}_\text{cl}\propto N_\text{u}^2$, see e.g. \cite{chao06}.

The Compton sources produce incoherent radiation. This statement follows from a simple estimation: Even for a long-period laser, $\lambda_\text{las}=10\,\mu\text{m}$, and a low-energy electrons, $\gamma = 100$, the number of coherently radiated periods would not exceed 10.

The introduced coherency criterion \eqref{eq:cohdef} does not work for number of pulses emitted in average by single electron less than unity, $\xi\le 1$.
Rigorously speaking, due to Poisson distribution of photons in the case $\xi\ll 1$ coherency -- modulation in angular distribution -- will be of order $\xi$: $P_2(\xi)/P_1(\xi)\approx\xi/2$, which is least of practical interest.

Increase in the undulator brightness, which is directly connected to the coherency of single--electron on-axis radiation, for higher energy of emitted photons requires enlarging the undulator spatial period together with increasing the electrons energy: $\lambda_\text{u}/\gamma = \text{const}$. In this case the energy of on-axis photons will increase in direct proportion to the energy of electrons. Tapering of the undulators are crucial for coherency of radiation, untapered undulators provide much less on-axis density of radiation.

\begin{acknowledgments}
We wish to acknowledge helpful and encouraging discussions with Prof. J. Urakawa, Prof. A. Potylitsyn and Dr. A. Opanasenko.
This work is partially supported by the Ministry of Education and Science of Ukraine, project No\,1-13-15.

\end{acknowledgments}


\begin{thebibliography}{25}%
\makeatletter
\providecommand \@ifxundefined [1]{%
 \@ifx{#1\undefined}
}%
\providecommand \@ifnum [1]{%
 \ifnum #1\expandafter \@firstoftwo
 \else \expandafter \@secondoftwo
 \fi
}%
\providecommand \@ifx [1]{%
 \ifx #1\expandafter \@firstoftwo
 \else \expandafter \@secondoftwo
 \fi
}%
\providecommand \natexlab [1]{#1}%
\providecommand \enquote  [1]{``#1''}%
\providecommand \bibnamefont  [1]{#1}%
\providecommand \bibfnamefont [1]{#1}%
\providecommand \citenamefont [1]{#1}%
\providecommand \href@noop [0]{\@secondoftwo}%
\providecommand \href [0]{\begingroup \@sanitize@url \@href}%
\providecommand \@href[1]{\@@startlink{#1}\@@href}%
\providecommand \@@href[1]{\endgroup#1\@@endlink}%
\providecommand \@sanitize@url [0]{\catcode `\\12\catcode `\$12\catcode
  `\&12\catcode `\#12\catcode `\^12\catcode `\_12\catcode `\%12\relax}%
\providecommand \@@startlink[1]{}%
\providecommand \@@endlink[0]{}%
\providecommand \url  [0]{\begingroup\@sanitize@url \@url }%
\providecommand \@url [1]{\endgroup\@href {#1}{\urlprefix }}%
\providecommand \urlprefix  [0]{URL }%
\providecommand \Eprint [0]{\href }%
\providecommand \doibase [0]{http://dx.doi.org/}%
\providecommand \selectlanguage [0]{\@gobble}%
\providecommand \bibinfo  [0]{\@secondoftwo}%
\providecommand \bibfield  [0]{\@secondoftwo}%
\providecommand \translation [1]{[#1]}%
\providecommand \BibitemOpen [0]{}%
\providecommand \bibitemStop [0]{}%
\providecommand \bibitemNoStop [0]{.\EOS\space}%
\providecommand \EOS [0]{\spacefactor3000\relax}%
\providecommand \BibitemShut  [1]{\csname bibitem#1\endcsname}%
\let\auto@bib@innerbib\@empty
\bibitem [{\citenamefont {Adolphsen}\ \emph {et~al.}(2013)\citenamefont
  {Adolphsen}, \citenamefont {Barone}, \citenamefont {Barish}, \citenamefont
  {Buesser} \emph {et~al.}}]{ilctdr}%
  \BibitemOpen
  \bibinfo {editor} {\bibfnamefont {C.}~\bibnamefont {Adolphsen}}, \bibinfo
  {editor} {\bibfnamefont {M.}~\bibnamefont {Barone}}, \bibinfo {editor}
  {\bibfnamefont {B.}~\bibnamefont {Barish}}, \bibinfo {editor} {\bibfnamefont
  {K.}~\bibnamefont {Buesser}},  \emph {et~al.},\ eds.,\ \href@noop {} {\emph
  {\bibinfo {title} {ILC Technical Design Report}}},\ Vol.\ \bibinfo {volume}
  {3 -- Accelerator}\ (\bibinfo  {publisher} {Linear Collider Project},\
  \bibinfo {year} {2013})\BibitemShut {NoStop}%
\bibitem [{\citenamefont {Lebrun}\ \emph {et~al.}(2012)\citenamefont {Lebrun},
  \citenamefont {Linssen}, \citenamefont {Lucaci-Timoce}, \citenamefont
  {Schulte}, \citenamefont {Simon}, \citenamefont {Stapnes}, \citenamefont
  {Toge}, \citenamefont {Weerts},\ and\ \citenamefont {Wells}}]{clicdr}%
  \BibitemOpen
  \bibinfo {editor} {\bibfnamefont {P.}~\bibnamefont {Lebrun}}, \bibinfo
  {editor} {\bibfnamefont {L.}~\bibnamefont {Linssen}}, \bibinfo {editor}
  {\bibfnamefont {A.}~\bibnamefont {Lucaci-Timoce}}, \bibinfo {editor}
  {\bibfnamefont {D.}~\bibnamefont {Schulte}}, \bibinfo {editor} {\bibfnamefont
  {F.}~\bibnamefont {Simon}}, \bibinfo {editor} {\bibfnamefont
  {S.}~\bibnamefont {Stapnes}}, \bibinfo {editor} {\bibfnamefont
  {N.}~\bibnamefont {Toge}}, \bibinfo {editor} {\bibfnamefont {H.}~\bibnamefont
  {Weerts}}, \ and\ \bibinfo {editor} {\bibfnamefont {J.}~\bibnamefont
  {Wells}},\ eds.,\ \href@noop {} {\emph {\bibinfo {title} {The CLIC Programme:
  towards a staged $e^+e^-$ Linear Collider exploring the Terascale, CLIC
  Conceptual Design Report}}}\ (\bibinfo  {publisher} {CERN-2012-005},\
  \bibinfo {year} {2012})\BibitemShut {NoStop}%
\bibitem [{\citenamefont {Kim}(1989)}]{kim88}%
  \BibitemOpen
  \bibfield  {author} {\bibinfo {author} {\bibfnamefont {K.-J.}\ \bibnamefont
  {Kim}},\ }in\ \href@noop {} {\emph {\bibinfo {booktitle} {AIP Conference
  Proceedings 184. Physics of Particle Accelerators. AIP New York}}}\ (\bibinfo
  {year} {1989})\ pp.\ \bibinfo {pages} {565--632}\BibitemShut {NoStop}%
\bibitem [{\citenamefont {Akhiezer}\ and\ \citenamefont
  {Shulga}(1996)}]{shulga96}%
  \BibitemOpen
  \bibfield  {author} {\bibinfo {author} {\bibfnamefont {A.}~\bibnamefont
  {Akhiezer}}\ and\ \bibinfo {author} {\bibfnamefont {N.}~\bibnamefont
  {Shulga}},\ }\href@noop {} {\emph {\bibinfo {title} {High--Energy
  Electrodynamics in Matter}}}\ (\bibinfo  {publisher} {Gordon and Breach},\
  \bibinfo {address} {Amsterdam},\ \bibinfo {year} {1996})\BibitemShut
  {NoStop}%
\bibitem [{\citenamefont {Hofman}(2004)}]{hofman04}%
  \BibitemOpen
  \bibfield  {author} {\bibinfo {author} {\bibfnamefont {A.}~\bibnamefont
  {Hofman}},\ }\href@noop {} {\emph {\bibinfo {title} {The physics of
  synchrotron radiation}}}\ (\bibinfo  {publisher} {Cambridge University
  Press},\ \bibinfo {address} {Cambridge, UK},\ \bibinfo {year}
  {2004})\BibitemShut {NoStop}%
\bibitem [{\citenamefont {Geloni}\ \emph {et~al.}(2012)\citenamefont {Geloni},
  \citenamefont {Kocharyan},\ and\ \citenamefont {Saldin}}]{geloni12}%
  \BibitemOpen
  \bibfield  {author} {\bibinfo {author} {\bibfnamefont {G.}~\bibnamefont
  {Geloni}}, \bibinfo {author} {\bibfnamefont {V.}~\bibnamefont {Kocharyan}}, \
  and\ \bibinfo {author} {\bibfnamefont {E.}~\bibnamefont {Saldin}},\
  }\href@noop {} {\enquote {\bibinfo {title} {On quantum effects in spontaneous
  emission by a relativistic electron beam in an undulator},}\ } (\bibinfo
  {year} {2012}),\ \Eprint {http://arxiv.org/abs/physics/1202.0691v1}
  {physics/1202.0691v1} \BibitemShut {NoStop}%
\bibitem [{\citenamefont {Klein}\ and\ \citenamefont {Nishina}(1929)}]{klein}%
  \BibitemOpen
  \bibfield  {author} {\bibinfo {author} {\bibfnamefont {O.}~\bibnamefont
  {Klein}}\ and\ \bibinfo {author} {\bibfnamefont {Y.}~\bibnamefont
  {Nishina}},\ }\href@noop {} {\bibfield  {journal} {\bibinfo  {journal} {Zs.
  f. Phys.}\ }\textbf {\bibinfo {volume} {52}},\ \bibinfo {pages} {853}
  (\bibinfo {year} {1929})}\BibitemShut {NoStop}%
\bibitem [{\citenamefont {Kincaid}(1977)}]{kincaid77}%
  \BibitemOpen
  \bibfield  {author} {\bibinfo {author} {\bibfnamefont {B.~M.}\ \bibnamefont
  {Kincaid}},\ }\href@noop {} {\bibfield  {journal} {\bibinfo  {journal}
  {Journal of Applied Physics}\ }\textbf {\bibinfo {volume} {48}},\ \bibinfo
  {pages} {2684} (\bibinfo {year} {1977})}\BibitemShut {NoStop}%
\bibitem [{\citenamefont {Howells}\ and\ \citenamefont
  {B.M.Kincaid}(1992)}]{howells92}%
  \BibitemOpen
  \bibfield  {author} {\bibinfo {author} {\bibfnamefont {M.}~\bibnamefont
  {Howells}}\ and\ \bibinfo {author} {\bibnamefont {B.M.Kincaid}},\ }\href@noop
  {} {\emph {\bibinfo {title} {The properties of undulator radiation}}},\
  \bibinfo {type} {Tech. Rep.}\ (\bibinfo  {institution} {LBL-34751, UC-406},\
  \bibinfo {year} {1992})\BibitemShut {NoStop}%
\bibitem [{\citenamefont {Bosco}\ and\ \citenamefont {Colson}(1983)}]{bosco83}%
  \BibitemOpen
  \bibfield  {author} {\bibinfo {author} {\bibfnamefont {P.}~\bibnamefont
  {Bosco}}\ and\ \bibinfo {author} {\bibfnamefont {W.~B.}\ \bibnamefont
  {Colson}},\ }\href@noop {} {\bibfield  {journal} {\bibinfo  {journal} {Phys.
  Rev. A}\ }\textbf {\bibinfo {volume} {28}},\ \bibinfo {pages} {319} (\bibinfo
  {year} {1983})}\BibitemShut {NoStop}%
\bibitem [{\citenamefont {Robb}\ and\ \citenamefont
  {Bonifacio}(2011)}]{robb11}%
  \BibitemOpen
  \bibfield  {author} {\bibinfo {author} {\bibfnamefont {G.}~\bibnamefont
  {Robb}}\ and\ \bibinfo {author} {\bibfnamefont {R.}~\bibnamefont
  {Bonifacio}},\ }\href@noop {} {\bibfield  {journal} {\bibinfo  {journal}
  {Europhysics Letters}\ }\textbf {\bibinfo {volume} {94}},\ \bibinfo {pages}
  {34002} (\bibinfo {year} {2011})}\BibitemShut {NoStop}%
\bibitem [{\citenamefont {Agapov}\ and\ \citenamefont
  {Geloni}(2014)}]{agapov14}%
  \BibitemOpen
  \bibfield  {author} {\bibinfo {author} {\bibfnamefont {I.}~\bibnamefont
  {Agapov}}\ and\ \bibinfo {author} {\bibfnamefont {G.}~\bibnamefont
  {Geloni}},\ }\href@noop {} {\bibfield  {journal} {\bibinfo  {journal} {Phys.
  Rev. ST Accel. Beams}\ }\textbf {\bibinfo {volume} {17}},\ \bibinfo {pages}
  {110704} (\bibinfo {year} {2014})}\BibitemShut {NoStop}%
\bibitem [{\citenamefont {Kolchuzhkin}\ \emph {et~al.}(2003)\citenamefont
  {Kolchuzhkin}, \citenamefont {Potylitsyn}, \citenamefont {Strokov},\ and\
  \citenamefont {Ababiy}}]{kolchuzhkin2003}%
  \BibitemOpen
  \bibfield  {author} {\bibinfo {author} {\bibfnamefont {A.}~\bibnamefont
  {Kolchuzhkin}}, \bibinfo {author} {\bibfnamefont {A.}~\bibnamefont
  {Potylitsyn}}, \bibinfo {author} {\bibfnamefont {S.}~\bibnamefont {Strokov}},
  \ and\ \bibinfo {author} {\bibfnamefont {V.}~\bibnamefont {Ababiy}},\
  }\href@noop {} {\bibfield  {journal} {\bibinfo  {journal} {Nucl. Instr. Meth.
  B}\ }\textbf {\bibinfo {volume} {201}},\ \bibinfo {pages} {307} (\bibinfo
  {year} {2003})}\BibitemShut {NoStop}%
\bibitem [{\citenamefont {Feller}(1957)}]{feller57}%
  \BibitemOpen
  \bibfield  {author} {\bibinfo {author} {\bibfnamefont {W.}~\bibnamefont
  {Feller}},\ }\href@noop {} {\emph {\bibinfo {title} {An Introduction to
  Probability Theory and Its Applications}}},\ \bibinfo {edition} {2nd}\ ed.,\
  Vol.~\bibinfo {volume} {1}\ (\bibinfo  {publisher} {John Wiley \& Sons, Inc;
  Chapman \& Hall, Limited},\ \bibinfo {address} {New York, London},\ \bibinfo
  {year} {1957})\BibitemShut {NoStop}%
\bibitem [{\citenamefont {van Kampen}(1998)}]{vankampen98}%
  \BibitemOpen
  \bibfield  {author} {\bibinfo {author} {\bibfnamefont {N.}~\bibnamefont {van
  Kampen}},\ }\href@noop {} {\emph {\bibinfo {title} {Stochastic Processes in
  Physics and Chemistry}}}\ (\bibinfo  {publisher} {Elsevier},\ \bibinfo
  {address} {London},\ \bibinfo {year} {1998})\BibitemShut {NoStop}%
\bibitem [{\citenamefont {Artru}(2014)}]{artru14}%
  \BibitemOpen
  \bibfield  {author} {\bibinfo {author} {\bibfnamefont {X.}~\bibnamefont
  {Artru}},\ }\href@noop {} {\enquote {\bibinfo {title} {Classical and quantum
  phenomenology in radiation by relativistic electrons in matter or in external
  fields},}\ } (\bibinfo {year} {2014}),\ \Eprint
  {http://arxiv.org/abs/1412.2061 [physics.acc-ph] 5 Dec 2014} {1412.2061
  [physics.acc-ph] 5 Dec 2014} \BibitemShut {NoStop}%
\bibitem [{\citenamefont {Bulyak}\ and\ \citenamefont
  {Urakawa}(2014)}]{bulyak14a}%
  \BibitemOpen
  \bibfield  {author} {\bibinfo {author} {\bibfnamefont {E.}~\bibnamefont
  {Bulyak}}\ and\ \bibinfo {author} {\bibfnamefont {J.}~\bibnamefont
  {Urakawa}},\ }\href@noop {} {\bibfield  {journal} {\bibinfo  {journal}
  {Journal of Physics: Conference Series}\ }\textbf {\bibinfo {volume} {517}},\
  \bibinfo {pages} {012001} (\bibinfo {year} {2014})}\BibitemShut {NoStop}%
\bibitem [{\citenamefont {Kim}(1986)}]{xray}%
  \BibitemOpen
  \bibfield  {author} {\bibinfo {author} {\bibfnamefont {K.-J.}\ \bibnamefont
  {Kim}},\ }in\ \href@noop {} {\emph {\bibinfo {booktitle} {X--ray data
  booklet}}}\ (\bibinfo  {publisher} {University of California},\ \bibinfo
  {address} {Berkeley},\ \bibinfo {year} {1986})\ pp.\ \bibinfo {pages}
  {14--20}\BibitemShut {NoStop}%
\bibitem [{\citenamefont {Bulyak}\ and\ \citenamefont
  {Skomorokhov}(2005)}]{bulyak05}%
  \BibitemOpen
  \bibfield  {author} {\bibinfo {author} {\bibfnamefont {E.}~\bibnamefont
  {Bulyak}}\ and\ \bibinfo {author} {\bibfnamefont {V.}~\bibnamefont
  {Skomorokhov}},\ }\href@noop {} {\bibfield  {journal} {\bibinfo  {journal}
  {Phys. Rev. {ST} Accel. Beams}\ }\textbf {\bibinfo {volume} {8}},\ \bibinfo
  {pages} {030703} (\bibinfo {year} {2005})}\BibitemShut {NoStop}%
\bibitem [{\citenamefont {Klauder}\ and\ \citenamefont
  {Sudarshan}(1968)}]{klauder68}%
  \BibitemOpen
  \bibfield  {author} {\bibinfo {author} {\bibfnamefont {J.~R.}\ \bibnamefont
  {Klauder}}\ and\ \bibinfo {author} {\bibfnamefont {E.~C.~G.}\ \bibnamefont
  {Sudarshan}},\ }\href@noop {} {\emph {\bibinfo {title} {Fundamentals of
  Quantum Optics}}}\ (\bibinfo  {publisher} {Benjamin},\ \bibinfo {address}
  {New York},\ \bibinfo {year} {1968})\BibitemShut {NoStop}%
\bibitem [{\citenamefont {Born}\ and\ \citenamefont {Wolf}(1970)}]{born70}%
  \BibitemOpen
  \bibfield  {author} {\bibinfo {author} {\bibfnamefont {M.}~\bibnamefont
  {Born}}\ and\ \bibinfo {author} {\bibfnamefont {E.}~\bibnamefont {Wolf}},\
  }\href@noop {} {\emph {\bibinfo {title} {Principles of Optics}}},\ \bibinfo
  {edition} {4th}\ ed.\ (\bibinfo  {publisher} {Pergamon Press},\ \bibinfo
  {address} {Oxford, London, Edinbourgh, New York, Toronto Sydney, Paris,
  Braunschweig},\ \bibinfo {year} {1970})\BibitemShut {NoStop}%
\bibitem [{\citenamefont {Brabec}\ and\ \citenamefont {Krausz}(2000)}]{brabec}%
  \BibitemOpen
  \bibfield  {author} {\bibinfo {author} {\bibfnamefont {T.}~\bibnamefont
  {Brabec}}\ and\ \bibinfo {author} {\bibfnamefont {F.}~\bibnamefont
  {Krausz}},\ }\href@noop {} {\bibfield  {journal} {\bibinfo  {journal} {Rev.
  Mod. Phys.}\ }\textbf {\bibinfo {volume} {72}},\ \bibinfo {pages} {545}
  (\bibinfo {year} {2000})}\BibitemShut {NoStop}%
\bibitem [{\citenamefont {Wang}\ \emph {et~al.}(2009)\citenamefont {Wang},
  \citenamefont {Freund}, \citenamefont {Harder}, \citenamefont {Miner},
  \citenamefont {Murphy}, \citenamefont {Qian}, \citenamefont {Shen},\ and\
  \citenamefont {Yang}}]{wang09}%
  \BibitemOpen
  \bibfield  {author} {\bibinfo {author} {\bibfnamefont {X.~J.}\ \bibnamefont
  {Wang}}, \bibinfo {author} {\bibfnamefont {H.~P.}\ \bibnamefont {Freund}},
  \bibinfo {author} {\bibfnamefont {D.}~\bibnamefont {Harder}}, \bibinfo
  {author} {\bibfnamefont {W.~H.}\ \bibnamefont {Miner}}, \bibinfo {author}
  {\bibfnamefont {J.~B.}\ \bibnamefont {Murphy}}, \bibinfo {author}
  {\bibfnamefont {H.}~\bibnamefont {Qian}}, \bibinfo {author} {\bibfnamefont
  {Y.}~\bibnamefont {Shen}}, \ and\ \bibinfo {author} {\bibfnamefont
  {X.}~\bibnamefont {Yang}},\ }\href@noop {} {\bibfield  {journal} {\bibinfo
  {journal} {Phys. Rev. Letters}\ }\textbf {\bibinfo {volume} {103}},\ \bibinfo
  {pages} {154801} (\bibinfo {year} {2009})}\BibitemShut {NoStop}%
\bibitem [{\citenamefont {Mun}\ \emph {et~al.}(2014)\citenamefont {Mun},
  \citenamefont {Jeong}, \citenamefont {Vinokurov}, \citenamefont {Lee},
  \citenamefont {Jang}, \citenamefont {Park}, \citenamefont {Jeon},\ and\
  \citenamefont {Shin}}]{mun14}%
  \BibitemOpen
  \bibfield  {author} {\bibinfo {author} {\bibfnamefont {J.}~\bibnamefont
  {Mun}}, \bibinfo {author} {\bibfnamefont {Y.~U.}\ \bibnamefont {Jeong}},
  \bibinfo {author} {\bibfnamefont {N.~A.}\ \bibnamefont {Vinokurov}}, \bibinfo
  {author} {\bibfnamefont {K.}~\bibnamefont {Lee}}, \bibinfo {author}
  {\bibfnamefont {K.-H.}\ \bibnamefont {Jang}}, \bibinfo {author}
  {\bibfnamefont {S.~H.}\ \bibnamefont {Park}}, \bibinfo {author}
  {\bibfnamefont {M.~Y.}\ \bibnamefont {Jeon}}, \ and\ \bibinfo {author}
  {\bibfnamefont {S.-I.}\ \bibnamefont {Shin}},\ }\href@noop {} {\bibfield
  {journal} {\bibinfo  {journal} {Phys. Rev. ST Accel. Beams}\ }\textbf
  {\bibinfo {volume} {17}},\ \bibinfo {pages} {080701} (\bibinfo {year}
  {2014})}\BibitemShut {NoStop}%
\bibitem [{\citenamefont {Chao}\ and\ \citenamefont {Tigner}(2006)}]{chao06}%
  \BibitemOpen
  \bibfield  {author} {\bibinfo {author} {\bibfnamefont {A.}~\bibnamefont
  {Chao}}\ and\ \bibinfo {author} {\bibfnamefont {M.}~\bibnamefont {Tigner}},\
  }\href@noop {} {\emph {\bibinfo {title} {Handbook of Acelerator Physics and
  Engineering}}},\ \bibinfo {edition} {3rd}\ ed.\ (\bibinfo  {publisher} {World
  Scientific},\ \bibinfo {address} {Singapore},\ \bibinfo {year}
  {2006})\BibitemShut {NoStop}%
\end{thebibliography}

\providecommand{\noopsort}[1]{}\providecommand{\singleletter}[1]{#1}%

\end{document}